\documentstyle[12pt]{article}
\newcommand{\vsl}{\mbox{$\not{\hspace{-.17em}v}$}}
\newcommand{\Dsl}{\mbox{$\not{\hspace{-.35em}D}$}}

\newcommand{\Dslp}{\mbox{${\Dsl}_{\Vert}$}}
\newcommand{\Dslo}{\mbox{${\Dsl}_{\hspace{-.2em}\bot}$}}

\newcommand{\Do}{\mbox{${D}_{\hspace{-.2em}\bot}$}}
\newcommand{\ao}{\mbox{${a}_{\hspace{-.17em}\bot}$}}
\newcommand{\po}{\mbox{${p}_{\hspace{-.2em}\bot}$}}
\newcommand{\ko}{\mbox{${k}_{\hspace{-.2em}\bot}$}}
\newcommand{\gamo}{\mbox{${\gamma}_{\hspace{-.2em}\bot}$}}

\newcommand{\ap}{\mbox{${a}_{\Vert}$}}
\newcommand{\Dvecl}{\mbox{${\stackrel{\leftarrow}{D}}$}}
\newcommand{\Dslovl}{\mbox{${\stackrel{\leftarrow}{\Dslo}}$}}
\begin{document}
\begin{titlepage}
\begin{flushright}
MZ-TH/93-33
\end{flushright}
\begin{center}
\begin{LARGE}
\em{Two Different Formulations of the Heavy Quark Effective Theory$^1$}
\end{LARGE}

\vspace{2cm}
\begin{large}
S. Balk$^2$, A. Ilakovac$^3$, J. G. K\"orner$^3$ and D. Pirjol$^2$
\vspace{1cm}
\newline
Johannes Gutenberg-Universit\"at, Institut f\"ur Physik,\\
Staudinger Weg 7, D-55099 Mainz, Germany
\end{large}
\end{center}

\vspace{2cm}
{\em Abstract:} We point out that there exist two different formulations of
the Heavy Quark Effective Theory (HQET). The one formulation of HQET was
mostly developed at Harvard and involves the use of the equation of motion
to eliminate the small components of the heavy quark field. The
second formulation, developed in Mainz, involves a series of
Foldy-Wouthuysen-type field transformations which diagonalizes
the heavy quark
Lagrangian in terms of an effective quark and antiquark sector. Starting at
O($1/m_Q^2$) the two formulations are different in that their effective
Lagrangians, their effective currents, and their effective wave functions
differ. However, when these three differences are properly taken into
account, the two alternative formulations lead to identical transition or
S-matrix elements. This is demonstrated in an explicit example at
O($1/m_Q^2$).
We point to an essential
difficulty of the Harvard HQET in that the Harvard effective fields
are not properly normalized starting at order $O(1/m_Q^2)$.
We provide explicit
higher order expressions for the effective fields and
the Lagrangian in the
Mainz approach, and write down an O($1/m_Q^3$) nonabelian version of the Pauli
equation for the heavy quark effective field.\\
\vspace{2cm}\\
\begin{footnotesize}
---------\\
$^1$ Invited talk given by J.G.K\"orner at the conference "Hadron Structure
93" in Banska Stiavnica (Slovakia) (to be published in the Proceedings) and
at the "XXVII International Symposium on the Theory of Elementary Particles"
in Wendisch-Rietz (Germany) (1993)\\
$^2$ Supported by the Graduiertenkolleg "Teilchenphysik", Mainz\\
$^3$ Supported in part by the BMFT, Federal Republic of Germany, under
contract 06MZ730
\end{footnotesize}
\end{titlepage}

Let us get straight into the heart of the matter of what we want to
discuss in this talk by stating that there exist two different formulations
of the Heavy Quark Effective Theory (HQET) which will be referred to as the
Harvard HQET \cite{Geo:90,Man:92} and the Mainz HQET \cite{Kor:91,DaM:93}.

The HQET that is most widely in use was mostly developed at Harvard
\cite{Geo:90,Man:92}.
It is for this reason that we shall refer to this version as the
Harvard HQET. In the Harvard approach one first extracts the mass phase
from the heavy quark field $\psi_Q(x)$ and then splits up the residual
field into its "large component" and "small component" pieces $h(x)$ and
$H(x)$, respectively. Accordingly one writes
\begin{equation}
\psi_Q(x)=e^{-im_Qv\cdot x}(h(x)+H(x))\label{eq:WFHar}
\end{equation}
where
\begin{equation}
\vsl h(x)=h(x)
\end{equation}
and
\begin{equation}
\vsl H(x)=-H(x)
\end{equation}
and where $v_\mu$ is the four-velocity of the heavy quark,
$v_\mu=p_\mu/m_Q$. Unfortunately the nomenclature in terms of the "large"
component field $h(x)$ and the "small" component field $H(x)$ has been
somewhat tangled up in the course of developing the Harvard HQET. In the
following we shall drop reference to the x-dependence in the fields.

In the next step one takes "one-half" of the full equation of motion
of QCD, i.e.
\begin{equation}
(i\Dsl-m_Q)\psi_Q=0 \label{eq:EOMF}
\end{equation}
by applying the small component projector $(1-\vsl)/2$ to
Eq. (\ref{eq:EOMF}):
\begin{equation}
(iv\cdot D+2m_Q)H=i\Dslo h \label{eq:EOMH}
\end{equation}
In Eq. (\ref{eq:EOMH}) we have introduced the four-transverse component of
the covariant derivative
\begin{equation}
\Do^{\mu}=D^{\mu}-v\cdot Dv^{\mu}
\end{equation}
where the four-transversality is defined w.r.t the velocity $v_\mu$.
Eq. (\ref{eq:EOMH}) can then be inverted to obtain
\begin{equation}
H=\frac{1}{(iv\cdot D+2m_Q)}i\Dslo h\label{eq:H}
\end{equation}
The small component field $H$ can be seen to be related to the large
component field $h$ through a geometric series in the inverse power of the
heavy mass. The Harvard approach consists in eliminating the small
component field out of the theory via Eq. (\ref{eq:H}).
Before we proceed any further there are two asides that we want
to embroider on.

First, there is the concept of the four-transversality used in
Eqs. (\ref{eq:EOMH}) and (\ref{eq:H}). Note that any four-vector
$a_\mu$ can be
split into its components transverse and parallel to a given four-velocity
$v_\mu$ according to
\begin{equation}
a^\mu=\ao^\mu+\ap^\mu :=(a^\mu-v\cdot a v^\mu)+(v\cdot a v^\mu)
\end{equation}
The concept of four-transversality is an important concept in the formalism
of HQET as one is frequently referring to rest-frame objects where
$v_\mu=(1,0,0,0)$ and where four-transverse vectors reduce to pure
three-vectors $\ao^\mu\rightarrow\vec{a}$. This concerns the covariant
derivative in Eq. (\ref{eq:H}) as well as the relative momenta
$\ko^\mu=k^\mu-k\cdot v v^\mu$ and spin operators
$\gamo^\mu=\gamma^\mu-\vsl v^\mu$ that are needed in the construction of
the spin wave-functions of HQET \cite{Hus:93}. Explicit reference to the
{\it four}-transversality is important in order not to get confused with
the usual notion of transversality which refers to a three-transversality
, i.e. $\vec{a}=\vec{a}-\vec{a}\cdot\hat{k} \hat{k}$ ($\hat{k}$
is a three-momentum of unit
length). In fact, the concept of four-transversality should be quite
familiar from QED when one is considering virtual photon exchange. In the
rest frame of the virtual photon $\vec{q}=0$ the conserved vector current
amplitude $T_\mu=\langle b|J_\mu|a\rangle$, with $q^\mu$ $T_\mu=0$, reduces to
a three-vector corresponding to the spin-one nature of the virtual photon.
When expanding the conserved amplitude $T_\mu$ in terms of covariants in a
general frame this has to be done in
terms of four-transverse covariants e.g.
$\po^\mu=p^\mu-\frac{p\cdot q}{q^2} q^\mu$, where the four-transversality
is defined w.r.t. the momentum $q_\mu$ of the photon.

The second aside concerns an alternative derivation of the Harvard HQET
that was
proposed in \cite{Man:92}. The authors of \cite{Man:92} employ functional
integration techniques to integrate out the small component field $H$ from
the functional action. They then arrive at the same relation
Eq. (\ref{eq:H}). That the equation of motion and functional integration
derivations of HQET are entirely equivalent may be appreciated by
considering the following simple example using the classical Lagrangian
$L(x,y)=-\frac{x^2}{2} +yx$. If one wishes one may view $x$ as being
related to the small component field $H$ and $y$ as been related to the
large component field $h$. Next consider the classical action
integral
\begin{equation}
\int dy\int_{-\infty}^{\infty} dx e^{L(x,y)}=
\int dy\int_{-\infty}^{\infty} dxe^{-\frac{1}{2} x^2+yx}
  \label{eq:PI}
\end{equation}
The x-integration in (\ref{eq:PI}) can be done by the usual "completion
of the square" trick, i.e. by writing
$-\frac{1}{2}x^2+yx=-\frac{(x-y)^2}{2}+\frac{y^2}{2}$
One then arrives at
\begin{equation}
\int dy\int_{-\infty}^{\infty} dx e^{L(x,y)}=
  \sqrt{2\pi}\int dy e^\frac{y^2}{2}
\end{equation}
After having integrated out the x-degree of freedom the new Lagrangian reads
$L(y)=y^2/2$. The crucial observation is that the same Lagrangian is
obtained by using the equation of motion
obtained by variation of the $L(x,y)$
w.r.t. the x-degree of freedom, i.e. $\partial L(x,y)/\partial x=0$. This
then gives the equation of motion $-x+y=0$. Substituting for x in the
original Lagrangian $L(x,y)$ one obtains the same new Lagrangian $L(y)$ as
before. We have chosen this simple illustration in order to make the point
that the equation of motion and functional integration derivation of the
Harvard HQET are entirely equivalent. This can be appreciated without
having to go into the technicalities of the full functional integration
approach presented in \cite{Man:92}.

Returning to Eq. (\ref{eq:H}) one writes
\begin{equation}
H=\frac{1}{2m_Q} \frac{1}{(1+\frac{iv\cdot D}{2m_Q})}i\Dslo h=
  \frac{1}{2m_Q}\left( 1-\frac{iv\cdot D}{2m_Q}
    +\frac{(iv\cdot D)^2}{4m_Q^2}+\cdots \right) i\Dslo h \label{eq:Hexp}
\end{equation}
As mentioned before one then eliminates the small component field H by
substituting Eq. (\ref{eq:Hexp}) into the effective Lagrangian and
current expressions.
This defines the Harvard HQET in terms of a $1/m_Q$ geometric type series
expansion of the effective Lagrangian and current in terms of
the large component field $h$ only.

In the Mainz HQET \cite{Kor:91,DaM:93} one uses a series of exponential
Foldy-Wouthuysen-type field transformations which yield an exponential type
series expansion for the effective Lagrangian and the effective current.

At this point we want to be rather suggestive and write down geometric and
exponential series expansions for a given small parameter $a$. One has :
\begin{equation}
\mbox{geometric series:}\qquad\frac{1}{1-ia}=1+ia-a^2\cdots
\end{equation}
\begin{equation}
\mbox{exponential series:}\qquad e^{ia}=1+ia-\frac{1}{2}a^2\cdots
\end{equation}
The series have been arranged such that they coincide in the
first order term. Although the two series
would be an oversimplified representation of the two
effective theories it still makes the point correctly: The Harvard and
Mainz HQET's are different in that their effective Lagrangians and currents
are different starting at $O(1/m_Q^2)$. However, despite the fact that the
respective effective Lagrangians and currents are different, one expects
the physics i.e. physical on shell matrix elements to be the same in both
theories, if they are calculated correctly. That these issues are of
immediate practical concern is being evidenced by the fact that $1/m_Q^2$
and even higher order corrections are presently being disscused in the
literature \cite{Fal:93,Lep:92}.

As a next step we want to explain in somewhat more detail how one derives
the Mainz HQET. To start with let us briefly retrace the physics steps that
lead one from the full QCD Lagrangian to the static heavy quark Lagrangian.
This will first be done in a completely heuristic manner. Consider again
the full QCD Lagrangian
\begin{equation}
{\cal L}=\bar{\psi}_Q(i\Dsl-m_Q)\psi_Q\label{eq:LQCD}
\end{equation}

The static approximation consists in neglecting the three-derivative in
Eq. (\ref{eq:LQCD}) relative to $m_Q$ according to the expansion
\begin{equation}
E=\sqrt{m_Q^2+\vec{p}\, ^2}
 = m_Q\left( 1+\frac{\vec{p}\, ^2}{2m_Q^2} +\cdots \right)
\end{equation}
This can be achieved by an appropriate field redefinition
$\psi_Q\rightarrow\psi'_Q$. This will then lead to
\begin{equation}
\mbox{\hspace*{-4cm}} \mbox{Step I:} \mbox{\hspace*{4cm}}
=\bar{\psi'}_Q(i\gamma_0 D_0-m_Q)\psi'_Q
\end{equation}
In the next step one shifts the energy scale $E\rightarrow E'=E-m_Q$ which
can again be achieved by an appropriate field redefinition
$\psi'_Q\rightarrow h_Q$. The Lagrangian now reads
\begin{equation}
\mbox{\hspace*{-4cm}} \mbox{Step II:} \mbox{\hspace*{4cm}}
=\bar{h}_Q(i\gamma_0 D_0)h_Q\label{eq:LHQT}
\end{equation}
If one chooses to work in terms of the heavy quark field $h^{(+)}$ only
(with $h^{(+)}=\frac{1}{2} (1+\gamma_0)h_Q$) one then recovers the static
Eichten-Hill Lagrangian ${\cal L}=\bar{h}^{(+)\dagger}iD_0 h^{(+)}$
\cite{Eic:90}.

The step-wise reduction of the QCD Lagrangian (\ref{eq:LQCD}) to the final form
(\ref{eq:LHQT}) can be achieved by a series of Foldy-Wouthuysen-type field
redefinitions which eventually yields the leading term result
(\ref{eq:LHQT}) as well as all higher dimension operators in the $1/m_Q$
expansion \cite{Hus:93}. The first transformation is $(j=1,2,3)$
\begin{eqnarray}
\mbox{\hspace*{-4cm}} \mbox{Step I:} \mbox{\hspace*{4cm}}  &&
\psi_Q\rightarrow e^{i\gamma_j\vec{D}_j/2m_Q}\psi'_Q\nonumber\\
&& \bar{\psi}_Q\rightarrow \bar{\psi'}_Q e^{-i\gamma_j\Dvecl_j/2m_Q}
\end{eqnarray}
where the arrow on the derivative indicates in which direction the
derivative acts. The heavy quark Lagrangian now becomes
\begin{equation}
{\cal L}=\bar{\psi'}_Q(i\gamma_0 D_0-m_Q)\psi'_Q
        +\sum_{k=1}^{\infty}\left( \frac{1}{m_Q} \right) ^k
         \bar{\psi'}_Q{\cal O}_k\psi'_Q\label{eq:Lexp}
\end{equation}
giving a form of the Lagrangian which makes explicit the mass perturbations.

The second transformation that removes the heavy mass dependence in the
first term of (\ref{eq:Lexp}) can be seen to be given by
\begin{eqnarray}
\mbox{\hspace*{-4cm}} \mbox{Step II:} \mbox{\hspace*{4cm}}  &&
 \psi_Q\rightarrow e^{-im_Q \gamma_0 t} h_Q\nonumber\\
&& \bar{\psi}_Q\rightarrow \bar{h}_Q e^{im_Q \gamma_0 t}\label{eq:psitr}
\end{eqnarray}

However, in order not to bring in further numerator mass terms through the
transformation (\ref{eq:psitr}) one needs to first block-diagonalize the
higher order operators ${\cal O}_k$ appearing in (\ref{eq:Lexp}). The
block-diagonalization
has to be done w.r.t. the upper and lower component of the
heavy quark fields $\frac{1+\gamma_0}{2} \psi_Q$ and
$\frac{1-\gamma_0}{2} \psi_Q$, respectively. This can be achieved order by
order by splitting the operator ${\cal O}_k$ into two pieces ${\cal O}_k^c$
and ${\cal O}_k^a$ that commute and anticommute with $\gamma_0$, respectively.
That is, one writes
\begin{equation}
{\cal O}_k={\cal O}_k^c+{\cal O}_k^a
\end{equation}
where
\begin{equation}
{\cal O}_k^{c,a}=\frac{1}{2}({\cal O}_k\pm\gamma_0{\cal O}_k\gamma_0)
\end{equation}
and $[{\cal O}_k^c,\gamma_0]=0$ and $\{ {\cal O}_k^a,\gamma_0\}=0$. The
anticommuting operator ${\cal O}_k^a$ is then removed from that order of
$(1/m_Q)^k$ by a further exponential type transformation. It is literally
shifted down to become part of the higher dimension operator
${\cal O}_{k+1}$.

Note that one never introduces any further implicit
time-derivatives $\partial_0$
through the field redefinitions in addition to the time-derivative in the
lowest order term of the final Lagrangian (see (\ref{eq:Lexp})).
Technically speaking, explicit
higher order time-derivative terms do appear
through the above field redefinitions.
However, these higher order
 time-derivatives always come in as commutators that are related
to the field strengh tensor (see Eq.(31))\cite{Kor:91,DaM:93}.
 This will
become important when we discuss the Pauli equation later on.

Up to now we have remained in the rest frame $v_\mu=(1,0,0,0)$ of the heavy
quark (or antiquark) in order to stay as close as possible to the heuristic
considerations that led us to the static Lagrangian (\ref{eq:LHQT}). The
field redefinitions that lead to the HQET Lagrangian can in fact be done in
any moving frame $v_\mu$ with $\vec{v}\neq 0$ by effecting the replacements
\begin{eqnarray}
\gamma_0 D_0\rightarrow \vsl v\cdot D:=\Dslp\nonumber\\
-\gamma_j D_j\rightarrow \Dsl-\vsl v\cdot D:=\Dslo
\end{eqnarray}

In the moving frame the heavy quark effective Lagrangian is then given by
\begin{eqnarray}\lefteqn{
{\cal L}_{HQET}^{KT}=\bar{h}_Q\{i\vsl v\cdot D
         +\frac{1}{2m_Q}(-D^2+(v\cdot D)^2   }\nonumber\\
& &      +\frac{i}{2}g\sigma_{\mu\nu}F^{\mu\nu}
         -ig\gamma_\mu\vsl F^{\mu\nu}v_\nu)+\cdots\} h_Q
         \label{eq:lkt}
\end{eqnarray}
where $F^{\mu\nu}$ is the field strength tensor
$F^{\mu\nu}=[D_\mu,D_\nu]$ ($D_\mu=\partial_\mu-igA^\mu$ and $A=A^a t^a$)
and where we have kept terms up to ${\cal O}(1/m_Q)$ only.
We chose to label the Mainz effective theory by
the initials of the authors of \cite{Kor:91}.
Note that the HQET
Lagrangian contains both heavy quark fields $h^{(+)}$ and heavy
antiquark fields $h^{(-)}$ where $\vsl h^{(\pm)}=\pm h^{(\pm)}$ and
$h_Q=h^{(+)}+h^{(-)}$. Remember that
this is different in the Harvard approach
 where the construction is done either in the quark sector or in the
antiquark sector by adjusting the sign of the mass phase in Eq.(1).
 The quark and antiquark sector in the Mainz HQET are completely
decoupled (at any order!) which is guaranteed by the very procedure of
deriving the HQET Lagrangian. To be sure one can easily convince oneself
that the nondiagonal contributions
$\bar h ^{(+)} \dots h ^{(-)}$ and $\bar h ^{(-)} \dots h ^{(+)}$
 induced by the first order terms
$\sigma_{\mu\nu} F^{\mu\nu}$ and $\gamma_\mu \vsl F^{\mu\nu}v_\nu$
in Eq.(\ref{eq:lkt}) cancel.
On the other hand, the $\gamma_\mu \vsl F^{\mu\nu}v_\nu$
contribution vanishes for the diagonal contributions
$\bar{h}^{(+)}\cdots h^{(+)}$ and $h^{(-)}\cdots h^{(-)}$
and one thereby
recovers the ${\cal O}(1/m_Q)$ Harvard HQET Lagrangian as promised
before (see Eq. (\ref{eq:Lhhar})).

As noted before the method of deriving the Mainz HQET proceeds step-wise
order by order. This is an iterative procedure which lends itself to
computer implementation. In fact we have written an efficient program in
Mainz that achieves just this. For the fun of it the program has been run
up to ${\cal O}(1/m_Q^{12})$ \cite{Ila:93}. As a sample result we list the
Mainz HQET Lagrangian for the heavy quark field $h^{(+)}$ up to order
${\cal O}(1/m_Q^5)$. One has
\begin{eqnarray}\lefteqn{
{\cal L_{HQET}^{KT}}=\bar{h}^{(+)}\left[ i\Dslp-\frac{1}{2m_Q}\Dslo^2
        -\frac{i}{4m_Q^2}\left( \frac{1}{2}\Dslp\Dslo^2+\Dslo\Dslp\Dslo
                         +\frac{1}{2}\Dslo^2\Dslp \right) \right.       }
                              \nonumber\\
& &     +\frac{1}{8m_Q^3}\left( \Dslp\Dslo\Dslp\Dslo+\Dslp\Dslo^2 \Dslp
                         +\Dslo\Dslp^2 \Dslo+\Dslo\Dslp\Dslo\Dslp
                         +\Dslo^4\right)
                              \nonumber\\
& &     +\frac{i}{16m_Q^4}\left( \frac{1}{2}\Dslp^2 \Dslo\Dslp\Dslo
                          +\frac{1}{2}\Dslp^2 \Dslo^2 \Dslp
                          +\frac{3}{2}\Dslp\Dslo\Dslp^2 \Dslo \right.
                              \nonumber\\
& &                       +2\Dslp\Dslo\Dslp\Dslo\Dslp
                          +\frac{1}{2}\Dslp\Dslo^2 \Dslp^2
                          +\frac{11}{8}\Dslp\Dslo^4
                          +\Dslo\Dslp^3 \Dslo
                               \nonumber\\
& &                       +\frac{3}{2}\Dslo\Dslp^2\Dslo\Dslp
                          +\frac{1}{2}\Dslo\Dslp\Dslo\Dslp^2
                          +\frac{3}{2}\Dslo\Dslp\Dslo^3
                               \nonumber\\
& &              \left.   +\frac{1}{4}\Dslo^2 \Dslp\Dslo^2
                          +\frac{3}{2}\Dslo^3 \Dslp\Dslo
                          +\frac{11}{8}\Dslo^4 \Dslp \right)
                               \nonumber\\
& &     -\frac{1}{32m_Q^5}\left( \Dslp^2 \Dslo\Dslp^2 \Dslo
                          +2\Dslp^2 \Dslo\Dslp\Dslo\Dslp
                          +\Dslp^2 \Dslo^2 \Dslp^2 \right.
                               \nonumber\\
& &                       +\frac{4}{3}\Dslp^2 \Dslo^4
                          +2\Dslp\Dslo\Dslp^3 \Dslo
                          +4\Dslp\Dslo\Dslp^2 \Dslo\Dslp
                               \nonumber\\
& &                       +2\Dslp\Dslo\Dslp\Dslo\Dslp^2
                          +\frac{10}{3}\Dslp\Dslo\Dslp\Dslo^3
                          +\Dslp\Dslo^2 \Dslp\Dslo^2
                               \nonumber\\
& &                       +\frac{5}{3}\Dslp\Dslo^3 \Dslp\Dslo
                          +\frac{4}{3}\Dslp\Dslo^4 \Dslp
                          +\Dslo\Dslp^4 \Dslo
                               \nonumber\\
& &                       +2\Dslo\Dslp^3 \Dslo\Dslp
                          +\Dslo\Dslp^2 \Dslo\Dslp^2
                          +2\Dslo\Dslp^2 \Dslo^3
                               \nonumber\\
& &                       +\Dslo\Dslp\Dslo\Dslp\Dslo^2
                          +2\Dslo\Dslp\Dslo^2 \Dslp\Dslo
                          +\frac{5}{3}\Dslo\Dslp\Dslo^3 \Dslp
                               \nonumber\\
& &                       +\Dslo^2 \Dslp\Dslo\Dslp\Dslo
                          +\Dslo^2 \Dslp\Dslo^2 \Dslp
                          +2\Dslo^3 \Dslp^2 \Dslo
                               \nonumber\\
& &    \left. \left.      +\frac{10}{3}\Dslo^3 \Dslp\Dslo\Dslp
                          +\frac{4}{3}\Dslo^4 \Dslp^2
                          +2\Dslo^6\right) \right] h^{(+)}
                            \label{eq:LHar}
\end{eqnarray}
As mentioned before, the time-derivative terms $\Dslp$ appearing at
$O(1/m_Q^2)$ and higher can all be rewritten in terms of the field strength
tensor \cite{Kor:91}.

The ensuing discussion will be in terms of the heavy quark field $h^{(+)}$
only and, for the sake of convenience, we shall omit the label $(+)$ on the
heavy quark field in the following.
In order to pinpoint the differences
in the Harvard and Mainz HQET's let us consider the heavy quark effective
Lagrangians and the flavour-conserving heavy quark effective currents up to
${\cal O}(1/m_Q^2)$. As mentioned before it is at this order where they start
to differ from one another. For the Harvard HQET one has
\begin{equation}
{\cal L}_{HQET}^{Harvard}=
        \bar{h}\{iv\cdot D-\frac{1}{2m_Q}\Dslo^2
       +\frac{i}{4m_Q^2}\Dslo v\cdot D\Dslo\} h \label{eq:Lhhar}\\
\end{equation}
\begin{eqnarray}\lefteqn{
J_{\mu,HQET}^{Harvard}=
        \bar{h}\{\Gamma_\mu
                +\frac{i}{2m_Q}(\Gamma_\mu\vec{\Dslo}-\Dslovl\Gamma_\mu)
                        }\nonumber\\
& &             +\frac{1}{4m_Q^2}(\Gamma_\mu v\cdot\vec{D}\vec{\Dslo}
                               +\Dslovl\Gamma_\mu\vec{\Dslo}
                               +\Dslovl v\cdot\Dvecl\Gamma_\mu)\}h
\end{eqnarray}
The ${\cal O}(1/m_Q^2)$ difference terms for the Mainz HQET are given by
\cite{Bal:93}
\begin{equation}
{\cal L}_{HQET}^{KT}={\cal L}_{HQET}^{Harvard}
                     +\frac{i}{4m_Q^2}\bar{h}
                      \left( -\frac{1}{2}\Dslo^2 v\cdot D
                          -\frac{1}{2}v\cdot D\Dslo^2
                      \right) h \label{eq:Ldif}\\
\end{equation}
\begin{eqnarray}
J_{\mu,HQET}^{KT}=J_{\mu,HQET}^{Harvard}
                 -\frac{1}{4m_Q^2}\bar{h}
                   \left( \frac{1}{2}\Dslovl^2\Gamma_\mu
                  +\frac{1}{2}\Gamma_\mu\vec{\Dslo}^2\right) h
                                    \label{eq:Jdif}
\end{eqnarray}
There are a number of observations we want to make about the difference terms
in Eqs. (\ref{eq:Ldif}) and (\ref{eq:Jdif}).

Let us first rewrite the difference term in the effective Lagrangian
(\ref{eq:Ldif}). One has
\begin{equation}
-\frac{1}{2}\Dslo^2 v\cdot D-\frac{1}{2}v\cdot D\Dslo^2
=-\Dslo v\cdot D\Dslo-\frac{1}{2}\Dslo[\Dslo,v\cdot D]
                     +\frac{1}{2}[\Dslo,v\cdot D]\Dslo\label{eq:Ldift}
\end{equation}
When one looks at the time derivative term $v\cdot D$
in (\ref{eq:Ldift}) (or alternatively
$D_0$ in the rest frame) one can see that the first term on the r.h.s. of
(\ref{eq:Ldift}) cancels the time derivative term in the Harvard Lagrangian
(\ref{eq:Lhhar}). The remaining two terms on the right hand side of
(\ref{eq:Ldif}) do not contain true time-derivatives since the commutator
can be expressed in terms of the field strength tensor via
\begin{equation}
[\Dslo,v\cdot D]=-igF^{\mu\nu}\gamma_\mu v_\nu
\end{equation}
The difference in the Mainz and Harvard Lagrangians lies in the fact that
the time-derivative terms only appear at leading order in the Mainz
Lagrangian but to all orders in the Harvard Lagrangian. This is exemplified
at second order in the above example. As a consequence of this one can
therefore quite easily write down Pauli equations to any desired order in
the Mainz approach by using the equation of motion for the effective heavy
quark field $h$ in the rest frame of the heavy quark. For example, this has
been done to ${\cal O}(1/m_Q^3)$ in \cite{Bal:93}. The result reads
\begin{eqnarray}\lefteqn{
i\frac{\partial h}{\partial t}=
   \left( gA^{0a}t^a+\frac{(\vec{P}-g\vec{A}^a t^a)^2}{2m}
   -\frac{g}{2m}\vec{\sigma}\cdot\vec{B}^a t^a
   -\frac{g}{8m^2}(\mbox{div}
  \vec{E}^a+f_{abc}\vec{A}^b\cdot\vec{E}^c)t^a \right.     }
       \nonumber\\
& &-\frac{ig}{8m^2}\vec{\sigma}\cdot \mbox{rot} \vec{E}^a t^a
   -\frac{ig^2}{8m^2}f_{abc}\vec{\sigma}\cdot(\vec{A}^b\times\vec{E}^c)t^a
   -\frac{g}{4m^2}\vec{\sigma}\cdot\vec{E}^a t^a\times(\vec{P}-g\vec{A}^b t^b)
       \nonumber\\
& &-\frac{1}{8m^3}[(\vec{P}-g\vec{A}^a t^a)^2
   -g\vec{\sigma}\cdot\vec{B}^a t^a]^2
       \nonumber\\
& & \left. +\frac{g^2}{8m^3}\left[ \vec{E}^a t^a\cdot \vec{E}^b t^b
         +\frac{i}{2}f_{abc}\vec{\sigma}\cdot
 (\vec{E}^a\times\vec{E}^b)t^c\right] \right) h
\end{eqnarray}
where $\vec{E}$ and $\vec{B}$ are the electric and magnetic colour
fields. To our
knowledge the Pauli equation has not been derived to this order before, let
alone in the non-Abelian case. We hope that we have convinced the reader by
now that the Mainz approach to HQET allows one to do so very efficiently.

As the next topic we want to discuss the calculation of physical matrix
elements using the two different formulations of HQET. We again concentrate
on the ${\cal O}(1/m_Q^2)$ contributions where the two formulations start to be
different.

We have arranged the difference terms in eq. (\ref{eq:Ldif}) and
(\ref{eq:Jdif}) in a rather suggestive manner by placing them one below
another: except for the $v\cdot D$ factors in the Lagrangian the difference
terms in the Lagrangian and in the current look quite similar to one
another. In fact one finds that the differences in the effective
Lagrangians "almost" cancel in physical transition matrix elements. They
cancel except for some ${\cal O}(1/m_Q^2)$ contributions which can
be absorbed in the
definition of the ${\cal O}(1/m_Q^2)$ HQET wave functions or the
interpolating field associated with the HQET state.

Let us explicitly
demonstrate this cancelation in the flavour-conserving
 case by using the Feynman
diagram language. One of the second order contributions of the Lagrangian
difference term  is drawn in Fig.1a. for $b\rightarrow b$ transitions.
The $v\cdot D$ term from the "nonlocal" Lagrangian insertion can be seen to
cancel against the inverse propagator $(v\cdot D)^{-1}$ adjoining it. In the
coordinate space language the two vertices in Fig.1a. are contracted to one
point due to the relation
$(iv\cdot D)\mbox{\hspace*{.5mm}}S(x,y)=\delta(x-y)$.
One thus remains
with the effective "local" insertion drawn in Fig.1b. This will be canceled
by the truly "local" $1/m_Q^2$ insertions from the effective current
difference (\ref{eq:Jdif}) which is not drawn here. Nonlocal insertions
in which $v\cdot D$
operates on the outer heavy quark wave function vanish due to the equation
of motion. There is one exception to this which is represented by the
Feynman diagram drawn in Fig.1c. The $v\cdot D$ to the left of the nonlocal
insertion is now canceled by the inverse propagator to the right of the
local insertion and one remains with a ${\cal O}(1/m_Q^2)$ difference even
for physical transitions. This difference corresponds to
a ${\cal O}(1/m_Q^2)$ renormalization of the HQET wave function or, more
precisely, the
interpolating field that is associated with the HQET state.
 The upshot of the
analysis is that the Mainz and Harvard formulations of HQET lead to
identical physical matrix elements as long as one takes into account the
differences in the ${\cal O}(1/m_Q^2)$ wave functions as drawn in Fig.2 in
addition to the differences in the effective Lagrangians and effective
currents. The same conclusion was reached in \cite{Bal:93} using functional
differentiation techniques.

However, in spite of the fact that the predictions for physical
matrix elements are identical, there exists an essential difference
between the two HQETs, in that the Harvard HQET lacks a correct
normalization of the heavy quark field.
We will illustrate the
problem in the context of QCD with unconfined heavy quarks.
When evaluating S-matrix elements for scattering processes by using the
usual machinery of the reduction formalism, an assumption is implicitly
made about the value of matrix elements like
\begin{equation}
    \langle 0|h(x)|Q(p,s)\rangle = \frac{1}{(2\pi)^{3/2}}
          \sqrt{\frac{m_Q}{E}}u(m_Q v,s)e^{-ip\cdot x}\,,
\end{equation}
giving the wave-function of the ingoing or outgoing heavy quarks.
This prejudice has its roots in the fact that such a relation holds
indeed true (by construction), if the heavy quark field in the effective
theory $h(x)$ is replaced by the QCD field $Q(x)$. However, this
relation {\em is not true} if $h(x)$ is the heavy quark field in the
Harvard HQET. The right-hand side has to be multiplied with a correction
factor of the form
\begin{equation}
     1 + \frac{p^2}{8m_Q^2} + \cdots
\end{equation}
which deviates from unity by an amount proportional to the residual
momentum $p$ of the one--quark state $|Q(p,s)\rangle$. It is clear that
the omission of this correction factor, such as it is done in the
naive application of the Harvard HQET, will possibly lead to a
wrong answer.

  On the other hand, relation (33) is valid to any order in the Mainz HQET.
This has been shown in \cite{Bal:93} by proving that the
heavy quark field in the Mainz effective theory is connected to the QCD field
by a unitary transformation. The fact that the Harvard HQET field has
a different normalization can be appreciated by noting that the two
are related by
\begin{equation}
h^{KT}(x) = \left( 1 + \frac{(\gamma\cdot D_{\perp})^2}{8m_Q^2}\right)
h^{Harvard}\,.
\end{equation}

  Another way of looking at the issue of the normalization of the heavy
quark field is to examine the form of the heavy quark propagator in the
vicinity of the one-particle pole. The requirement of correct normalization
can be expressed by saying that the residue of this pole should be
unity. That these two points of view are related can be seen from the
form of the K\"all\'en--Lehmann representation for the heavy quark
propagator \cite{BjDr:68}
\begin{eqnarray}
   S'_F(p) = \frac{1+\gamma\cdot v}{2} \int_0^{\infty}\mbox{d}M^2
            \frac{\rho(M^2)}{i(v\cdot p - M^2 + i\epsilon)}
\end{eqnarray}
where $\rho(M^2)$ is defined by a sum over all possible intermediate
states
\begin{equation}
 \rho(p^2)\delta_{\alpha\beta} =
(2\pi)^3\sum_n\delta^{(4)} (p_n-p)\langle 0|h_{\alpha}(0)|n\rangle
                                  \langle n|\bar h_{\beta}(0)|0\rangle\,.
\end{equation}
If one then makes use of Eq.(33), the contribution of the one-quark
intermediate state to the spectral function $\rho(M^2)$ can be
seen to be as follows
\begin{eqnarray}
\rho(M^2) &=& \delta(M^2)\,,
\end{eqnarray}
which gives a free field-like behaviour of the heavy quark propagator
in the neighbourhood of
the one-particle pole. On the other hand, in the Harvard HQET
the residue of the propagator at the one-particle pole takes a different
value (which is even a function of the residual momentum of the heavy quark).
This has been shown through explicit calculation in a recent paper by
A.Das \cite{Das:93}.

The lack of maintaining the correct normalization of the field or wave
functions by eliminating the "small" component via the "equation of motion"
approach has long been known in the Abelian context of QED, where these
issues were studied in connection with the Foldy-Wouthuysen transformation
\cite{Fey:61,Pau:58,Fuk:54,Oku:54}. Eliminating the "small" component naively
leads to a nonhermitean Hamiltonian. In the case of an electron interacting
with an external electric field, the lowest order manifestation of the
non-hermiticity is an imaginary electric dipole moment. All these issues
were very nicely discussed in the paper by A. Das \cite{Das:93}.\\
\vspace{2ex}\\
{\em Acknowledgement}: We owe many thanks to George Thompson who
generously shared with us his profound insights into Quantum Field Theory.
He had always emphasized that the two methods of deriving the Harvard HQET
discussed in the text are equivalent and, for illustration, provided us
with the simple classical analogue presented in the text.

\newpage
.\newline
\begin{Large}
{\em Figure Captions}
\vspace{.5cm}
\end{Large}\\
{\em Fig.1} $O(1/m_Q^2)$ contributions from the effective Lagrangian to the
flavour-conserving
 $H_b (v)\rightarrow H_b (v')$ current transition. a) and b) show how
the nonlocal contribution a) becomes a local current contribution b)
through the contraction of two interaction points. In c) we show an
external line contribution which can be absorbed into the definition of the
HQET wave function.\\ \vspace{.3cm}
{\em Fig.2} $O(1/m_Q^2)$ difference of Harvard and Mainz HQET wave
functions.
\end{document}